\documentclass[preprint,aps]{revtex4}

\usepackage{amsmath,amssymb,amsfonts,dcolumn,color,graphicx,graphics,setspace,latexsym,setspace,lscape,subfigure,placeins,epsfig,hyperref}

\newcommand{\be}{\begin{equation}}
\newcommand{\ee}{\end{equation}}
\newcommand{\ba}{\begin{eqnarray}}
\newcommand{\ea}{\end{eqnarray}}

\newcommand{\lb}{\left}
\newcommand{\rb}{\right}

\begin{document}

\title{\Large \bf Ricci flow of unwarped and warped product manifolds}

\author{Sanjit Das, Kartik Prabhu and Sayan Kar}
\email{sanjit@cts.iitkgp.ernet.in, kartikprabhu.iitkgp@gmail.com, sayan@cts.iitkgp.ernet.in}
\affiliation{\rm Department of Physics and Meteorology {\it and} Center for Theoretical Studies \\Indian Institute of Technology, Kharagpur, 721302, India}

\begin{abstract}

We analyse Ricci flow (normalised/un-normalised) of product manifolds
--unwarped as well as warped, through a study of generic examples.
First, we investigate such flows for the unwarped scenario with
manifolds of the type $\mathbb S^n\times \mathbb S^m$, $\mathbb S^n\times \mathbb H^m$, $\mathbb H^m\times \mathbb H^n$
and also, similar multiple products. We are able to single out
generic features such as singularity formation, isotropisation at
particular values of the flow parameter and evolution characteristics.
Subsequently, motivated by warped braneworlds and extra dimensions,
we look at Ricci flows of warped spacetimes. Here, we are able to
find analytic solutions for a special case by variable separation. For others we numerically
solve the equations and draw certain useful inferences about
the evolution of the warp factor, the scalar curvature as well
the occurrence of singularities at finite values of the flow parameter. We 
also investigate the dependence of the singularities of the flow on the 
initial conditions. We expect our results to be useful in any 
physical/mathematical context 
where such product manifolds may arise.
\end{abstract}


\maketitle

\section{Introduction and overview} 
Ricci and other geometric flows {\cite{ricci}} have been a topic of
active research interest in both mathematics and physics, over the
last few years. Though introduced many years ago by Hamilton \cite{
hamilton} in mathematics and in the work of Friedan \cite{friedan} in the
context of $\sigma$-models and string theory,
it was Perelman's proof of the Poincare conjecture using Ricci flows
\cite{perelman} which generated most of the current interest. Much of the 
subsequent work has
thus focused on using ideas associated with Ricci flows, in the context
of general relativity and gravitation \cite{gr}. In this article, we
follow a similar line of thought with reference to metrics on 
warped manifolds of various types.  

The definition of a warped manifold in mathematics is somewhat different
from that used in physics. In mathematics, product manifolds with different
overall conformal factors dependent only on a certain parameter, are termed
as warped. However, in physics, warping refers to an extra coordinate
dependence of one or more of these conformal factors--in other words, 
the conformal
factor associated with one manifold in the product when assumed to
be dependent on one or more of the coordinates of the other manifold, 
is said to be warped.
Otherwise, it is termed as unwarped. We shall follow this latter
definition in our work here.

We restrict our attention to Ricci flow of specific types of product manifolds.
Earlier work on such manifolds can be found in \cite{maxu}.
We begin, following the work on non--Bianchi types in the penultimate
section in \cite{isenberg},
by looking at products of the form
$\mathbb S^n\times \mathbb S^m$, $\mathbb S^n\times \mathbb H^m$, $\mathbb H^m\times \mathbb H^n$ for arbitrary $m$ and $n$, and with different
conformal factors (dependent on a parameter $\lambda$) associated with
the canonical metric on each manifold in the product. This study essentially
involves unwarped products in the sense described in the previous
paragraph. We also look at multiple products (more than two) as
a possible generalisation. What do we look for in our investigations?
Apart from exact solutions of the flow equations (which may not be
possible always), 
we focus attention on the following aspects: (a)
blow-up/dropping to zero value of any of the conformal factors at a 
finite $\lambda$, (b) isotropisation (i.e. equal value of the conformal
factor at a finite $\lambda$) (c) fixed points/curves of the flows 
(d) dependence of flow on initial conditions.
   
In the second half of our paper, we
look at specific warped product manifolds largely inspired by
currently fashionable ideas in the physics of extra dimensions
. In particular, our product manifolds here are essentially
similar to those used extensively in warped braneworld models
introduced in the work of Randall and Sundrum, about a decade ago
\cite{brane}. A Ricci flow of such manifolds seems to provide
interesting pointers in the physics of braneworlds--an aspect
we discuss towards the end of our article.

We have studied normalised Ricci flows in the unwarped cases
while for the warped ones we primarily investigate the unnormalised Ricci 
flow. In the former, we are mainly concerned with Riemannian metrics,
whereas in the latter we discuss the Lorentzian case in detail because
it is the one which is physically relevant.

A word here about the essential mathematical motivation of Ricci
flows may not be too inappropriate.
In any method to find the `best' metric on a manifold the first thing
to do it to define a natural evolution 
of metrics. One such evolution of metrics w.r.t. a `time' parameter (not the
physical time, in any sense) is the Ricci flow.  
Once we have defined the flow it is important to prove that the flow 
exists for all time and converges to a geometric limit. In a case the flow 
does not converge, the corresponding metrics degenerate and one then 
needs to relate the degeneration with the topology of the manifold.
Among many good candidates, a specific choice is
an evolution equation 
of vector fields (rather, a family of vector fields) on the space of metrics
known as the Ricci flow equation and defined as: 
\begin{equation}
\frac{\partial g_{ij}}{\partial \lambda}=-2R_{ij}
\end{equation}
For the normalised Ricci flow, the corresponding equation turns
out to be:
\begin{equation}
\frac{\partial g_{ij}}{\partial \lambda}=-2R_{ij} + \frac{2}{n}\langle R\rangle
g_{ij}
\end{equation}
where $\langle R\rangle =\frac{\int R dV}{\int dV}$. Thus the normalised
flow ceases to be different from the unnormalised one if we consider
non--compact, infinite volume manifolds (with a finite value for $\int R dV$).
Moreover, for constant curvature manifolds the second term in the normalised
flow equation reduces to $\frac{2}{n} R$.
In this article, we shall first look at normalised flows in the next section.
Subsequently, we shall deal with un--normalised flows.
\section{Ricci flow of unwarped product manifolds}
As mentioned in the Introduction, in this Section, we focus attention
on unwarped product manifolds. We discuss each case separately
by studying the evolution equations and solving each dynamical
system analytically. We present our results through figures which
illustrate the nature of evolution of each conformal factor in
the corresponding product manifold. In this Section, our study is
based on normalised Ricci flows, since all our manifolds are of
constant curvature.
\subsection{On $\mathbb S^n(\sqrt D)\times \mathbb S^m(\sqrt E)$}
Here we focus on product manifolds of the aforementioned type 
with $D$ and $E$ representing the conformal factors for the
$\mathbb S^n$ and $\mathbb S^m$ in the product. $D$ and $E$ are
functions of the flow parameter $\lambda$.

Let us compute the various curvature quantities(Ricci tensor, Ricci scalar) 
of this product manifold. We denote the metric on 
$\mathbb S^n(\sqrt a)$ by $a g_{\mathbb S^n}$ where $ g_{\mathbb S^n}$ is 
the canonical metric on $n$ -sphere. In the same way, 
the metric on the product manifold $\mathbb S^n(\sqrt D)\times \mathbb S^m(\sqrt E)$ looks like:
$g=D g_{\mathbb S^n} + E g_{\mathbb S^m}$.
Let $Y$ and $V$ be the unit vector fields on $\mathbb S^n$ and $\mathbb S^m$ 
respectively and $X$ be a unit vector field either on $\mathbb S^n$ or 
$\mathbb S^m$ which is perpendicular to both $Y$ and $V$. It can be shown 
that (by Koszul's formula) $\langle\nabla_Y X,V\rangle =0$ as $[Y,X]$ is either zero or 
tangent to $\mathbb S^n$ and likewise with $[X,V]$. Thus $\nabla_Y X=0$ if 
$X$ is tangent to $\mathbb S^m$. Also $\nabla_Y X$ is tangent to  
$\mathbb S^n$ if $X$ is tangent to  $\mathbb S^n$, showing that $\nabla_Y X$ can be computed on  $\mathbb S^n(\sqrt D)$. This shows that if $X$, $Y$ are tangent to $\mathbb S^n$ and $U$, $V$ are  tangent to $\mathbb S^m(\sqrt E)$, then 
 $\mathfrak R(X \wedge V)=0$ where  $\mathfrak R$ is the curvature operator. 
Also  $\mathfrak R(X \wedge Y)=\frac{1}{D}X \wedge Y$ and $\mathfrak R(U \wedge V)=\frac{1}{E}U \wedge V$. In particular, all sectional curvatures lie in 
the interval $[0,max\{\frac{1}{D},\frac{1}{E}\}]$. From this we can say: 
$Rc  (X)=\frac {(n-1)}{D} X$\\
$Rc  (V)=\frac {(m-1)}{E} V$ or we can write $Rc_g=(n-1)  g_{\mathbb S^n}+(m-1) g_{\mathbb S^m}$ and
$R=\frac {n(n-1)}{D}+\frac {m(m-1)}{E}$. Therefore we can make inference 
that $\mathbb S^n(\sqrt D)\times \mathbb S^m(\sqrt E)$ always has constant 
scalar curvature and it is an Einstein manifold exactly when 
$\frac{n-1}{D}=\frac{m-1}{E}$ (this is possible either $n$, $m$ $\ge 2$ 
or $n=m=1$). It is also worthwhile to mention that this product manifold 
has constant sectional curvature only when $n=m=1$. 
Now we have all the ingredients in our hand to write normalised Ricci flow 
equations. These are given as:
\begin{eqnarray}
\frac{d D}{d \lambda}=-2(n-1)+\frac{2}{(n+m)}\left \{\frac{n(n-1)}{D}+\frac{m(m-1)}{E}\right \}D \\ 
\frac{d E}{d \lambda}=-2(m-1)+\frac{2}{(n+m)}\left \{\frac{n(n-1)}{D}+\frac{m(m-1)}{E}\right \}E
\end{eqnarray}
The above two equations represent a first order dynamical system. For different
$n$ and $m$ we get different dynamical systems. For $n,m=1$ we have $D$ and
$E$ as constant and the metric is a flat metric on $\mathbb S^1\times \mathbb S^1$. 
\subsubsection{General Solution}
After some simplification, we have:
\begin{eqnarray}
\frac{d D}{d\lambda} =\,-\frac{2m(n-1)}{m+n}+\frac{2m(m-1)}{(n+m)}\,\frac{D}{E}\\
\frac{d E}{d\lambda}=\,-\frac{2n(m-1)}{m+n}+\frac{2n(n-1)}{(n+m)}\,\frac{E}{D}
\end{eqnarray}

We may analyse the following general form for the above pair of
equations. Similar equations will arise for
the other product manifolds to be discussed later on. 
Solutions can be found for arbitrary  $a$, $b$, $c$ and $d$.
Specific values can be inserted for each case separately.
Thus, we now analyse the generic system:
\begin{eqnarray}
\frac{d x}{d \lambda}=-a+b~\frac{x}{y}\hspace{0.2in};\hspace{0.2in}
\frac{d y}{d \lambda}=-c+d~\frac{y}{x}
\end{eqnarray}

Dividing one equation with the other, we get:
\begin{equation}
\frac{dx}{dy}=\frac{y(yd-cx)}{x(bx-ay)}
\end{equation} 
which can be solved by putting $y=vx$:
\begin{equation}
x \frac{dv}{dx}=\frac{v^2(a+d)-v(c+b)}{(b-av)}\equiv \frac{v^2 c_1+v c_2}{(b-av)}
\end{equation}
Defining $c_1=\,(d+a)$ and $c_2=\,-(c+b)$,
the general solution of $x$ and $v$ will look like:
\begin{equation}
\frac{b}{c_2}\ln(\frac{v}{c_1 v+c_2})-\frac{a}{c_1}\ln(c_1 v+c_2)=\,\ln x - \ln k
\end{equation}
For $\mathbb S^n(\sqrt D)\times \mathbb S^m(\sqrt E)$, with $c_1=2(n-1)$
and $c_2=-2(m-1)$ we have:
\begin{equation}
\ln x+\frac{m}{m+n}\,\ln v=\,\ln k, \Longrightarrow   x^n\,y^m=\,k
\end{equation}
If we put the above back in the original equations, we get the profile of 
$x$ or $y$.
\begin{equation}
\frac{d x}{d \lambda}=\,-a+\frac{b}{k}\,x^{\frac{m+n}{n}}
\end{equation}
This can be integrated to get $\lambda(x)$ in terms of hypergeometric functions.
For $n=m$ or $n=1,m$, as we shall see one can obtain $x$ and $y$ as 
functions of $\lambda$. Otherwise, in general, the functions $\lambda (x)$
or $\lambda(y)$ are not invertible.
We now discuss a few illustrative, special cases.

\subsubsection{Special cases}

{{\bf a.} $\mathbb S^1(\sqrt D)\times \mathbb S^2(\sqrt E)$}\\

\begin{figure}
\centering
\subfigure[$D_0 = E_0= 1 $  for $\mathbb S^1 \times \mathbb S^2$]{\includegraphics[width = 0.4\textwidth]{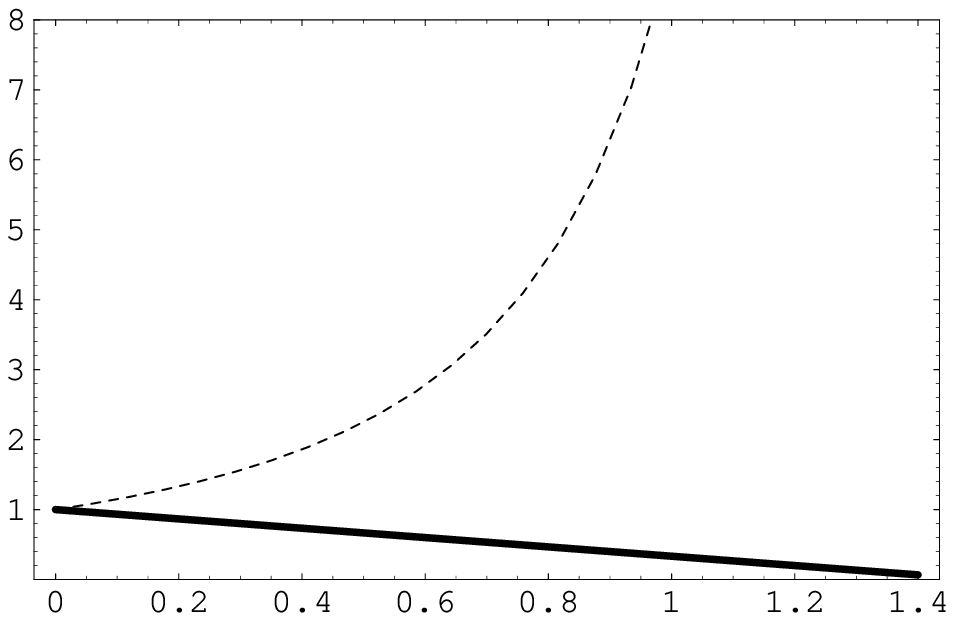}\label{rffig1}}
\subfigure[$D_0 =2$, $ E_0=3$ for $\mathbb S^2 \times \mathbb S^3$]{\includegraphics[width = 0.4\textwidth]{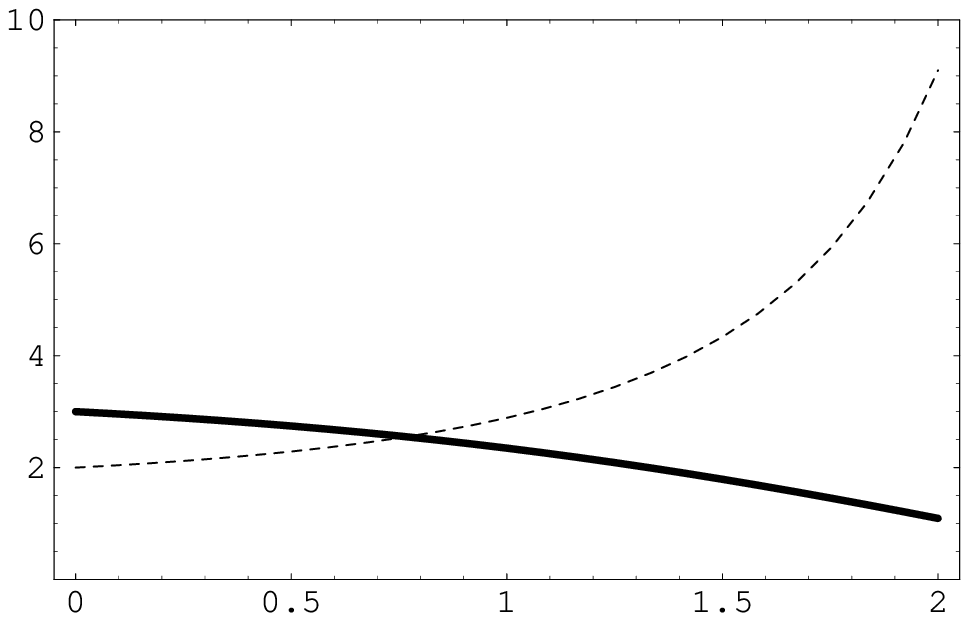}\label{rffig4}}
\subfigure[$D_0=1$, $E_0=0.9999$ for $\mathbb S^2 \times \mathbb S^2$]{\includegraphics[width = 0.4\textwidth]{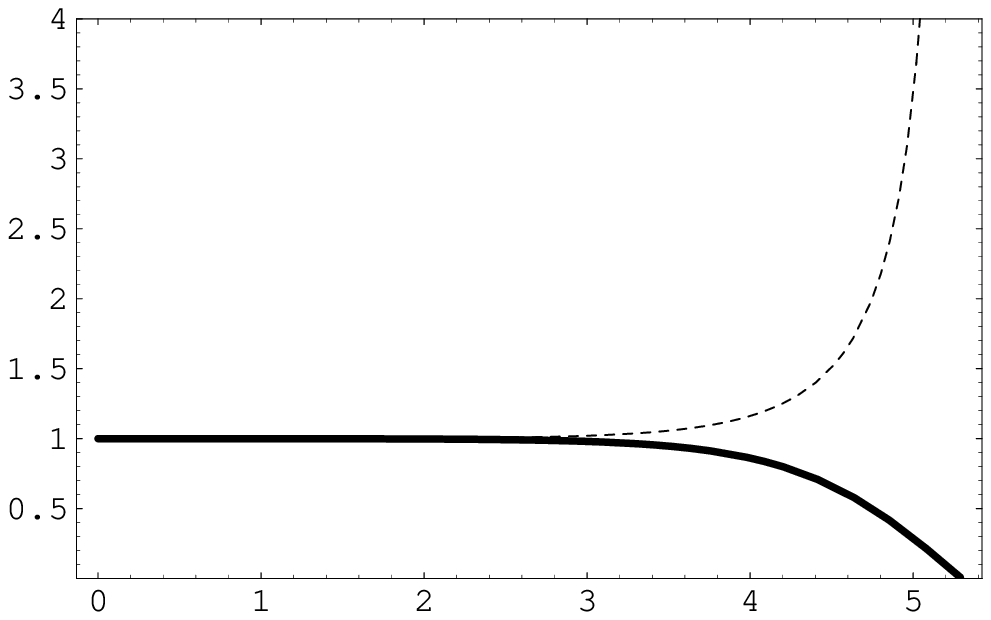}\label{rffig2}}
\subfigure[$D_0=0.9999$, $E_0=1$ for $\mathbb S^2 \times \mathbb S^2$]{\includegraphics[width = 0.4\textwidth]{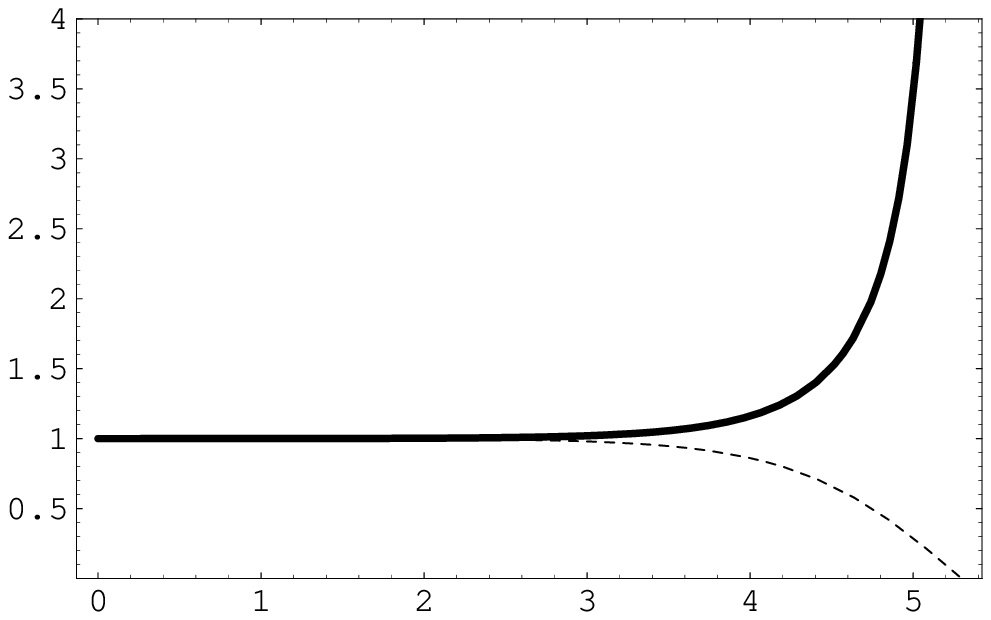}\label{rffig3}}
\caption{$D(\lambda)$ (dashed) and $E (\lambda)$ (continuous). The horizontal axes is $\lambda$.}
\end{figure}

Here $n=1$ and $m=2$. The equations become:
\begin{eqnarray}
\frac{d D}{d \lambda}=\frac{4}{3}\frac{D}{E}\hspace{0.1in};\hspace{0.1in}
\frac{d E}{d \lambda}=-\frac{2}{3}
\end{eqnarray}
The solutions are straightforward:
\begin{equation}
D(\lambda)= D_0 \left (\frac{E_0}{E_0-\frac{2}{3}\lambda}\right )^2 \hspace{0.2in};\hspace{0.2in} E (\lambda)=E_0-\frac{2}{3}\lambda
\end{equation}
where at $\lambda=0$, $D=D_0, E=E_0$.
One notices that D increases with increasing $\lambda$ and diverges
to infinity at a finite $\lambda=\frac{3}{2}E_0$. At the same value of
$\lambda$, E drops to zero. The flow therefore ends at this value of
$\lambda$--a curvature singularity appearing because of the shrinking of
the $S^1$ part to zero radius. Fig.\ref{rffig1} shows the aforementioned behaviour. \\

In general, for products of the type $\mathbb S^1(\sqrt D)\times \mathbb S^m(\sqrt E)$
we end up with the following expressions for $D(\lambda)$ and $E(\lambda)$.

\begin{equation}
D(\lambda)= D_0 \left (\frac{E_0}{E_0-\frac{2(m-1)}{(m+1)}\lambda}\right )^m \hspace{0.2in};\hspace{0.2in} E (\lambda)=E_0-\frac{2 (m-1)}{(m+1)}\lambda
\end{equation}
The conclusions for arbitrary $m$, are similar to those for $m=2$. \\[10pt]

{{\bf b.} $\mathbb S^2(\sqrt D)\times \mathbb S^2(\sqrt E)$}\\

Here, we turn to the more interesting case of $\mathbb S^2(\sqrt D)\times \mathbb S^2(\sqrt E)$. The equations are:
\begin{eqnarray}
\frac{d D}{d \lambda}=-1+\frac{D}{E}\hspace{0.1in} ; \hspace{0.1in}
\frac{d E}{d \lambda}=-1+\frac{E}{D}
\end{eqnarray}
We easily note that $ DE=constant=c(say)$. 
Putting $\frac{1}{E}=\frac{D}{c}$ the first equation looks like:
\begin{equation}
\frac{dD}{d\lambda}=\frac{D^2-c}{c}
\end{equation}
 and the solution with, $D(\lambda=0)=D_0$ is:
\begin{equation}
D(\lambda)=\sqrt c\Big[\frac{D_0(1+\exp^{\frac{2\lambda}{\sqrt c}})+\sqrt c(1-\exp^{\frac{2\lambda}{\sqrt c}})}{D_0(1-\exp^{\frac{2\lambda}{\sqrt c}})+\sqrt c(1+\exp^{\frac{2\lambda}{\sqrt c}})}\Big]
\end{equation}
From $DE=c$ one can find $E(\lambda)$ which is given as:
\begin{equation}
E(\lambda)=\sqrt c\left[\frac{D_0(1-\exp^{\frac{2\lambda}{\sqrt c}})+\sqrt c(1+\exp^{\frac{2\lambda}{\sqrt c}})}{D_0(1+\exp^{\frac{2\lambda}{\sqrt c}})+\sqrt c(1-\exp^{\frac{2\lambda}{\sqrt c}})}\right ]
\end{equation}
Thus, it is clear that $E_0=\frac{c}{D_0}$. For $D_0=E_0=\sqrt{c}$, we get
$D(\lambda)$ and $E(\lambda)$ as constant for all $\lambda$. For $D_0 >\sqrt{c}$
we have a divergence in $D(\lambda)$ while $E(\lambda)$ drops to zero at the
same $\lambda=\lambda_d$. On the other hand, if $D_0<\sqrt{c}$, the behaviour
of $D(\lambda)$ and $E(\lambda)$ are reversed. In either case the
value of $\lambda$ at which the divergence occurs is given as:
\begin{equation}
\lambda_d = \frac{\sqrt c}{2}\ln \left (\frac{D_0+\sqrt c}{D_0-\sqrt c}\right )
\end{equation}
Fig.\ref{rffig2} and \ref{rffig3} demonstrate these features.

It may be noted that the above results for $\mathbb S^2(\sqrt D)\times \mathbb S^2(\sqrt E)$ holds in the more general setting, 
for $\mathbb S^n(\sqrt D)\times \mathbb S^n(\sqrt E)$
with an appropriate scaling of $\lambda$. The general equations are:
\begin{eqnarray}
\frac{d D}{d \lambda}=(n-1)(-1+\frac{D}{E}) \hspace{0.1in;\hspace{0.1in}
\frac{d E}{d \lambda}=(n-1)(-1+\frac{E}{D})
\end{eqnarray}
The solution for $D(\lambda)$ turns out to be:
\begin{equation}
D(\lambda)=\sqrt c\Big[\frac{D_0(1+\exp^{\frac{2(n-1)\lambda}{\sqrt c}})+\sqrt c(1-\exp^{\frac{2(n-1)\lambda}{\sqrt c}})}{D_0(1-\exp^{\frac{2(n-1)\lambda}{\sqrt c}})+\sqrt c(1+\exp^{\frac{2(n-1)\lambda}{\sqrt c}})}\Big]
\end{equation}
Further $\lambda_d$ is given as:
\begin{equation}
\lambda_d=\frac{\sqrt c}{2(n-1)}\ln(\frac{D_0+\sqrt c}{D_0-\sqrt c})}
\end{equation}
It is worthwhile to note that in this case, $D=E=$constant represents
a fixed point (soliton) of the flow. \\[20pt]

\subsubsection{Numerical Evaluations}
Though we have exact solutions formally, it is easier to analyse the
nature of the flow if we obtain numerical solutions of the dynamical systems
involved. This is because the analytic solutions is the cases other
than the ones quoted above are necessarily non--invertible. \\ 

We shall focus here on cases where $n\neq m$. 
Let us see if it is possible to have $D(\lambda_0)= E(\lambda_0)$
for $n\neq m$. Looking at the general equations one can evaluate
the derivatives assuming this ($D=E$) to be true. It turns out that,
\begin{eqnarray}
{\lb[\frac{dD}{d\lambda}\rb]}_{\lambda_0} = \frac{2m(m-n)}{n+m} \hspace{0.2in};\hspace{0.2in} 
{\lb[\frac{dE}{d\lambda}\rb]}_{\lambda_0} = \frac{2n(n-m)}{n+m}
\end{eqnarray}
Thus for $n=m$ there is no chance of D to be equal to E at some $\lambda$.
On the other hand, this is always possible for $n\neq m$. 
We illustrate this fact by solving the equations numerically,
as a dynamical system.\\[10pt]

{{\bf a.} $\mathbb S^2(\sqrt D)\times \mathbb S^3(\sqrt E)$}\\

The equations in this case are:
\begin{eqnarray}
\frac{d D}{d \lambda}=-\frac{6}{5}+\frac{12}{5}\frac{D}{E} \hspace{0.2in};\hspace{0.2in}
\frac{d E}{d \lambda}=-\frac{8}{5}+\frac{4}{5}\frac{E}{D}
\end{eqnarray}
Fig.\ref{rffig4} below demonstrates the facts mentioned above.

Similar results may be obtained for any other case with $n\neq m$.
In summary, we observe the following:

\noindent $\bullet$ There are fixed curves defined by $\frac{D}{E}=\frac{n-1}{m-1}$.

\noindent $\bullet$ As one of the radii (D or E) increases (or becomes singular), the
other drops to zero. This will also be apparent from a $\ln D$ vs $\ln E$ plot, which
will follow the equation:\\
 $\ln D = -\frac{m}{n}\ln E + \frac{1}{n} \ln k$. 

\noindent $\bullet$ Reversing the initial conditions can lead to a reversal of the
behaviour of D or E. This is easily seen in the case $n= m$ where the
equations go over to each other if we swap $D$ and $E$. However, this is
also true, as we have checked (not shown here) for $n\neq m$. 

\noindent $\bullet$ Generically similar behaviour for all $m$,
is also seen when one of the
two manifolds is $\mathbb S^1$ and the other is $\mathbb S^m$ (with $m\neq 1$).

\subsection{On $\mathbb S^n(\sqrt(D))\times \mathbb H^m(\sqrt(E))$}
Let us now turn towards analysing another type of product manifold
(as mentioned above). For this case,
the metric looks like:
$g=D g_{\mathbb S^n} + E g_{\mathbb H^m}$ 
where $g_{\mathbb S^n}$ is the canonical metric on $n$ dimensional sphere
and $g_{\mathbb H^m}$ is the metric on a m dimensional hyperbolic space.
The normalised Ricci flow equations turn out to be:
\begin{eqnarray}
\frac{d D}{d\lambda}=-2(n-1)+\frac{2}{(n+m)}\left \{\frac{n(n-1)}{D}-\frac{m(m-1)}{E}\right \}D \\
\frac{d E}{d \lambda}=2(m-1)+\frac{2}{(n+m)}\left \{\frac{n(n-1)}{D}-\frac{m(m-1)}{E}\right \}E
\end{eqnarray}
where, as before, $D$ and $E$ are functions of the flow parameter $\lambda$.

\subsubsection{ General Solution of this system}
As in the previous subsection, after some simplification we have:
\begin{eqnarray}
\dot{x}=\,-\frac{2m(n-1)}{m+n}-\frac{2m(m-1)}{(n+m)}\,\frac{x}{y} \hspace{0.2in};\hspace{0.2in}
\dot{y}=\,\frac{2n(m-1)}{m+n}+\frac{2n(n-1)}{(n+m)}\,\frac{y}{x}
\end{eqnarray}
Following the general analysis in terms of $x$, $y$ and the $a,b,c,d$
given earlier,  
 we note that the values of $a,b,c,d$, 
for this case are:
\begin{equation}
a=2m\frac{(n-1)}{(m+n)} \hspace{0.1in};\hspace{0.1in} 
b=-2m\frac{(m-1)}{(m+n)} \hspace{0.1in};\hspace{0.1in} 
c=-2n\frac{(m-1)}{(m+n)} \hspace{0.1in};\hspace{0.1in} 
d=2n\frac{(n-1)}{(m+n)}
\end{equation}
Also, $c_1=\,(d+a)=\,2(n-1), c_2=\,-(c+b)=\,2(m-1)$. Notice the
difference in signs as compared to the $\mathbb S^n \times \mathbb S^m$ case.  
Further, the relation between $x$ and $y$ will be:
\begin{equation}
\ln x+\frac{m}{m+n}\,\ln v=\,\ln k \Longrightarrow   x^n\,y^m=\,k
\end{equation}
Though this appears to be the same as for $\mathbb S^n \times \mathbb S^m$,
a difference will appear in the explicit solution of $x(\lambda)$ and 
$y(\lambda)$. For instance,
\begin{equation}
\frac{d x}{d \lambda}=\,-a-\frac{b}{k}\,x^{\frac{m+n}{n}}
\end{equation}
and, consequently, one can find $y(\lambda)$.
Let us look at some special cases now.
\subsubsection{$\mathbb S^1(\sqrt D)\times \mathbb H^2(\sqrt E)$}
In this rather simple case, the difference quoted above is easily seen.
The equations are:
\begin{equation}
\frac{d D}{d \lambda}=-\frac{4}{3}\frac{D}{E} \hspace{0.2in};\hspace{0.2in}
\frac{d E}{d \lambda}=\frac{2}{3} 
\end{equation}
The solution is:
\begin{equation}
D(\lambda)= D_0 \left (\frac{E_0}{E_0+\frac{2}{3}\lambda}\right )^2 \hspace{0.2in};\hspace{0.2in} E (\lambda)=E_0+\frac{2}{3}\lambda
\end{equation}
where at $\lambda=0$, $D=D_0, E=E_0$.
Note the change in sign, which suggests that there is no singularity
in the evolution of D, starting from $\lambda=0$, unlike the previous
case of $\mathbb S^1\times \mathbb S^2$. The result here can easily be generalised for $\mathbb S^1\times \mathbb H^m$. 

\subsubsection{$\mathbb S^n(\sqrt D)\times \mathbb H^n(\sqrt E)$}
Here, the equations turn out to be:
\begin{eqnarray}
\frac{d D}{d \lambda}=-(n-1)-(n-1)\frac{D}{E} \hspace{0.2in};\hspace{0.2in}
\frac{d E}{d \lambda}= (n-1) + (n-1)\frac{E}{D}
\end{eqnarray}
The features to note here are:

\noindent $\bullet$ There are no fixed curves, unlike the previous case
for $\mathbb S^m \times \mathbb S^m$.

\noindent $\bullet$ There is always a possibility of having $D=E$ at some
$\lambda$.

\noindent $\bullet$ As before $DE$ is a constant, but evolution of $D(\lambda)$
and $E(\lambda)$ differs as compared to the previous cases.

\subsection{$\mathbb H^n \times \mathbb H^m$}
Finally we turn to products of the form $\mathbb H^n \times \mathbb H^m$.
Here, the metric takes the form
$g=D g_{\mathbb H^n} + E g_{\mathbb H^m}$ 
where $g_{\mathbb H^n (m)}$ is the canonical metric on $n (m)$ dimensional 
hyperbolic space.
The normalised Ricci flow equations become:
\begin{eqnarray}
\frac{d D}{d \lambda}= 2(n-1)-\frac{2}{n+m}\left \{\frac{n(n-1)}{D}+\frac{m(m-1)}{E}\right \}D \\
\frac{d E}{d \lambda}= 2(m-1)-\frac{2}{n+m}\left \{\frac{n(n-1)}{D}+\frac{m(m-1)}{E}\right \}E
\end{eqnarray}
The analysis for these equations follows in the same way as for
the earlier cases. We list below are few characteristic observations.

\noindent $\bullet$ The fixed curves are given as $ \frac{D}{E}=\frac{n-1}{m-1}$--this is similar to the $\mathbb S^n \times \mathbb S^m$ case.

\noindent $\bullet$ The possibility of $D=E$ at some $\lambda$ exists as long as
$n\neq m$.

\noindent $\bullet$ The flow ends at singular points, depending on initial
conditions.

\subsection{Multiple products and general trends}
What could happen if we take multiple products of the $\mathbb S^n$, or $\mathbb H^n$
or mixed products of various kinds? In general, it is difficult
question to address. However, we can surely try out some
specific examples. We illustrate things using one such example, below.
\subsubsection{$\mathbb S^p(\sqrt F)\times \mathbb S^n(\sqrt D)\times \mathbb S^m(\sqrt E)$}
For this triple product, the metric looks like:
$g=F g_{\mathbb S^p}+D g_{\mathbb S^n} + E g_{\mathbb S^m}$ 
where $g_{\mathbb S^n}$ is the canonical metric on $n$ dimensional sphere.
The normalised Ricci flow equations are as follows:
\begin{equation}
 \frac{d F}{d \lambda}=-2(p-1)+\frac{2}{(p+n+m)}\left \{\frac{p(p-1)}{F}+ \frac{n(n-1)}{D}+\frac{m(m-1)}{E}\right \}F 
\end{equation}
\begin{equation}
 \frac{d D}{d\lambda}=-2(n-1)+\frac{2}{(p+n+m)}\left \{\frac{p(p-1)}{F}+ \frac{n(n-1)}{D}+\frac{m(m-1)}{E}\right \}D 
\end{equation}
\begin{equation}
\frac{d E}{d \lambda}=-2(m-1)+\frac{2}{(p+n+m)}\left \{\frac{p(p-1)}{F}+ \frac{n(n-1)}{D}+\frac{m(m-1)}{E}\right \}E 
\end{equation}
The above flow equations can be analysed in a way similar to the
cases discussed earlier. The dynamical system is far more complicated.
A few general comments can be made without solving the equations.

$\bullet$ If $n=m=p=1$ then F, D and E do not evolve.

$\bullet$ If any one of n,m or p is equal to one (say $p=1$), then
the system of equations for D, E can be solved independently as was
done for the simple products discussed in the previous subsections.
These solutions can be used to find the evolution of F. 

$\bullet$ If $n=m=p$ then the equations are simpler, but unlike the
case of a product of two manifolds, we do not have any fixed curves
here. If $n\neq m\neq p$ or $n=m\neq p$ there are fixed curves
which can be found by just putting the R. H. S. of the above
equations to zero and solving the algebraic equations.

$\bullet$  An intersection of all three functions
D, E and F at some $\lambda$ is possible as long as
$p<n<m$, $n^2+m^2>p(n+m)$ and $p^2+m^2<n(p+m)$. This can be checked by using
$D=E=F$ on the R. H. S. of the general equations and
assuming that $\frac{dF}{d\lambda}>0,\frac{dD}{d\lambda}<0$ at the same
$\lambda$. One can easily show that the above restrictions are
impossible to achieve. This is also visible in the numerical
example given below.

\subsubsection{$\mathbb S^2(\sqrt F)\times \mathbb S^3(\sqrt D) \times \mathbb S^4 (\sqrt{E})$}
\begin{figure}
\centering
\subfigure[$F_0=3$, $D_0=4$, $E_0=5$]{\includegraphics[width = 0.4\textwidth]{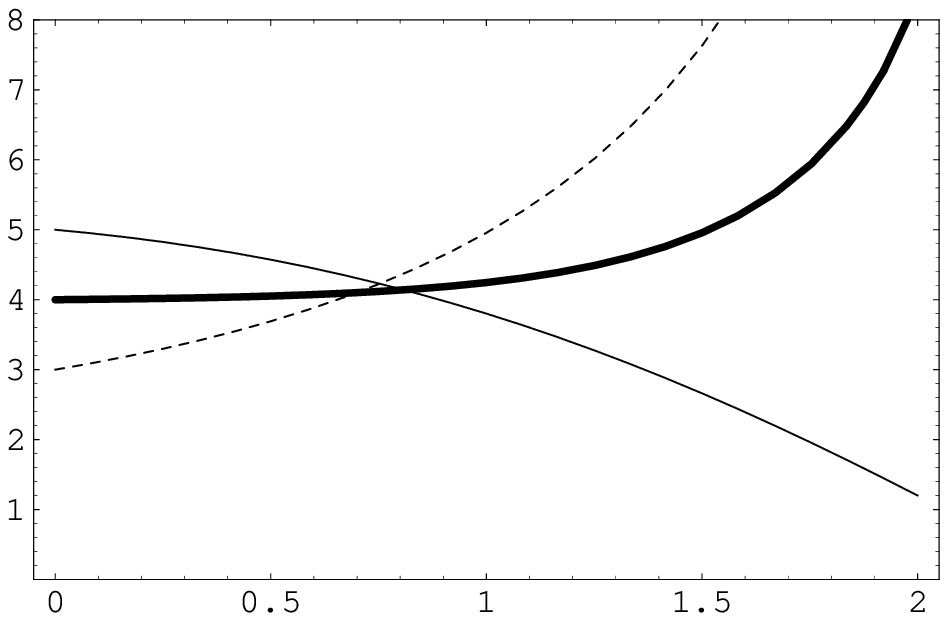}}
\subfigure[$F_0=3$, $D_0=2$, $E_0=3$]{\includegraphics[width = 0.4\textwidth]{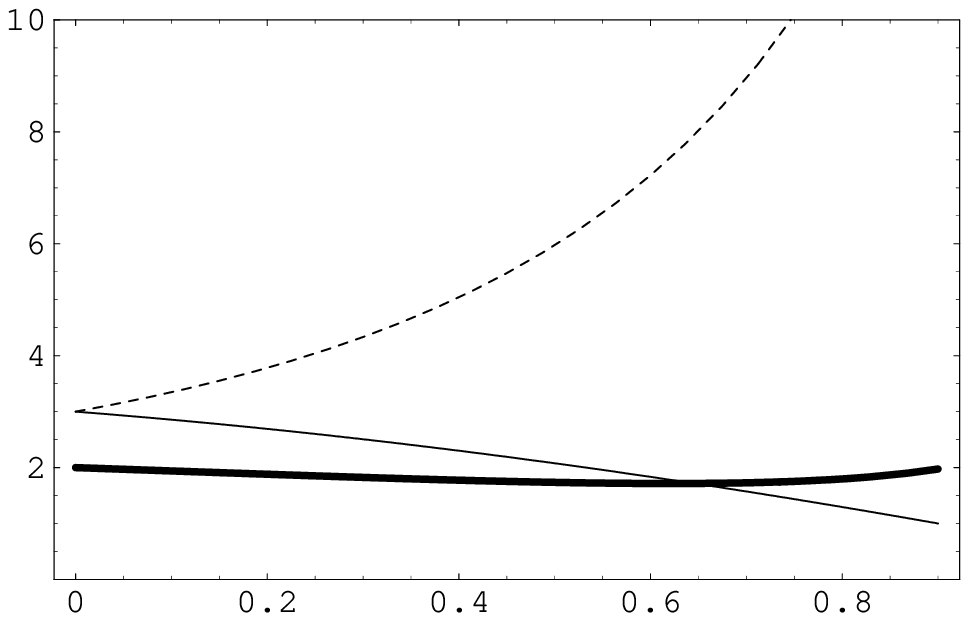}}
\caption{$F(\lambda)$ (dashed), $D(\lambda)$ (thick) and $E(\lambda)$ (thin) for $\mathbb S^2\times \mathbb S^3\times \mathbb S^4$. The horizontal axes is $\lambda$ }
\label{rffig56}
\end{figure}

In this case the equations are:
\begin{eqnarray}
\frac{dF}{d\lambda} = -\frac{14}{9} +\frac{4}{3} \frac{F}{D} + \frac{8}{3} \frac{F}{E} \hspace{0.2in};\hspace{0.2in} 
\frac{dD}{d\lambda} = -\frac{8}{3} +\frac{4}{9} \frac{D}{F} + \frac{8}{3} \frac{D}{E}
\end{eqnarray}
\begin{equation}
\frac{dE}{d\lambda} = -\frac{10}{3} +\frac{4}{9} \frac{E}{D} + \frac{4}{3} \frac{E}{F} 
\end{equation}
We numerically solve this system. Fig.\ref{rffig56} shows some typical evolution features for F, D and E.

The figures illustrate the fact that there can be a pair among $D$, $E$ and $F$
which can intersect at some $\lambda$. Different pairs intersect at
different points. Also, there can also be a situation without any intersection,
this seems to be the case when a pair among D. E and F have the same 
initial value. Furthermore, divergences appear and the
flow ends at a finite $\lambda$, as was observed in the earlier cases
with a product of two manifolds.\\

It is possible to generalise the above results to products of
arbitrary number of manifolds of differing dimensions. One can
easily write down the general evolution equations by inspection.
The case with products of $\mathbb S^m$ and $\mathbb H^m$ and
also only $\mathbb H^m$ can also discussed. Generic features 
are possibly not very different. 
Furthermore, instead of taking the normalisation on the whole
manifold, one may consider, following related work in \cite{isenberg},
cases with volume normalisation
along sections. We will discuss some of these issues in a later
article.
For the moment, we prefer to
switch over and move on to the perhaps more interesting case of
warped products.

\section{Ricci flow of warped product manifolds}

In spacetime models with extra dimensions,
braneworld scenarios \cite{brane} provide many examples of product 
manifolds. We consider one such case where the manifold is a product 
of Minkowski spacetime $\mathcal{M}$ and the real line $\mathcal{R}^1$.
In addition, we have warping in the sense described earlier in the Introduction,
which makes the ensuing analysis different from the ones described in the
previous sections.
 
In order to consider the evolution under Ricci flow of the 
\emph{warped product} $\mathcal{M} \times \mathcal{R}^1$ 
we assume the line element of the form:
			\be\label{metric}
			ds^2 = e^{2 f(\sigma,\lambda)} (-dt^2+dx^2+dy^2+dz^2) + r_c^2(\sigma, \lambda)d\sigma^2
			\ee
Note that the functions $f$ and $r_c$ are assumed to be functions of
$\sigma$ as well as a non--coordinate parameter $\lambda$. 
We now evolve the metric functions in the above line element
according to a un--normalised Ricci flow.
		with $\lambda$ as the flow parameter.
			
	Th relevant geometric quantities for this metric are --
			\be
			-R_{tt} = R_{xx} = R_{yy} = R_{zz} = -\frac{e^{2f}}{r_c^2}\left(f''+4f'^2-\frac{f'r_c'}{r_c}\right)
			\ee
			\be
			R_{\sigma \sigma} = -4 (f''+f'^2-\frac{f'r_c'}{r_c})
			\ee
			\be\label{ricci scalar}
			R = - \frac{4}{r_c^2}\left(2f''+5f'^2-2\frac{f'r_c'}{r_c}\right)
			\ee

After some straightforward algebra, the Ricci flow equations become --
			\be\label{f flow}
			\dot{f} = \frac{1}{r_c^2}\left(f'' + 4f'^2-\frac{f'r_c'}{r_c}\right)
			\ee
			\be\label{r flow}
			\dot{r_c} = \frac{4}{r_c}\left(f'' + f'^2-\frac{f'r_c'}{r_c}\right)
			\ee
		here ``   $\dot{f}$  '' represents $\frac{\partial f}{\partial\lambda}$ and ``  $f'$  '' represents $\frac{\partial f}{\partial \sigma}$ and similarly for $r_c$. The equation for $f$ may be thought
of as a certain type of \emph{nonlinear heat equation with the 
equation for $r_c$ appearing as a constraint}.

We make note of the remarkable fact that, the above Ricci flow equations 
are independent the signature of the 
part of the metric that contains a scaling with $e^{2f}$. 	
	
We now try to understand the behaviour of $f$ (especially as a
function of $\lambda$) by solving the flow equations. The resulting
 solutions will tell us how the metric, as well other geometric
quantities, evolve along the flow.
\subsection{Separable metric functions: exact results}
Let us first assume that 
\begin{equation}
r_c(\sigma, \lambda) = r_c(\lambda) \hspace{0.1in};\hspace{0.1in} 
f(\sigma ,\lambda) = f_\sigma (\sigma) + f_\lambda (\lambda)
\end{equation}
This assumption enables us to write the Eq.\ref{f flow} and Eq.\ref{r flow} 
in a variable separable form as:
			\be\label{f flow separable}
			r_c^2\dot{f_\lambda} = \left(f_\sigma '' + 4f_\sigma '^2\right)=K_1
			\ee
			\be\label{r flow separable}
			r_c\dot{r_c} = 4 \left(f_\sigma '' + f_\sigma '^2\right) = -K_2
			\ee\\
		For the $\sigma$-part we get --
			\be\label{f' f''}
 			f_\sigma ' = \pm \left(\frac{K_1}{3}+\frac{K_2}{12}\right)^{\frac{1}{2}} \hspace{5pt};\hspace{5pt} f_\sigma '' = - \left(\frac{K_1+K_2}{3}\right)
			\ee
			which imply $K_1 = -K_2$ and --
			\be\label{f sigma soln}
 			f_\sigma = \pm \left(\frac{K_1}{4}\right)^{1/2} \sigma
			\ee\\
		Therefore, we can easily solve the $\lambda$ evolution 
to get,
			\be\label{f r lambda}
 			r_c^2 = e^{2f_\lambda} = 1+2K_1\lambda
			\ee
		where, we make use of $K_1=-K_2$ and the initial condition 
that at $\lambda=0$ we have $f=0$ and $r_c = 1$ that is no scaling in the 
metric.

		To simplify, we can further write $K_1=1/2\lambda_c$ 
and thus the final metric turns out to be:
			\be\label{metric soln}
 			ds^2 = \left(1+\frac{\lambda}{\lambda_c}\right)\left[ \exp\left(\pm\frac{\sigma}{\sqrt{2\lambda_c}}\right) (-dt^2+dx^2+dy^2+dz^2) + d\sigma^2 \right]
			\ee
The Ricci scalar for the abovestated metric evolves as,
			\be\label{ricci soln}
 			R = -\frac{5/2}{\lambda+\lambda_c}	
			\ee
We can prove that the solution obtained above is the only solution of the 
set Eq.\ref{f flow} and Eq.\ref{r flow} when $r_c = r_c(\lambda)$. 
Under these assumptions Eq.\ref{r flow} can be written as:
			\be\label{B defn}
			r_c\dot{r_c} = 4 \left(f'' + f'^2\right) = 4 B(\lambda)^2
			\ee
		This readily yields $r_c^2 = 8\int B(\lambda)^2d\lambda = \beta$; and then solving for $f$,
			\[
			f_1 = \pm B(\lambda)\sigma + C(\lambda)
			\]
		or
			\[
			f_2 = A(\lambda) + \ln\left[ \cosh\left( B(\lambda)\sigma + C(\lambda) \right) \right]
			\]

		Using $f=f_1$ in Eq.\ref{f flow}, then gives --
			\[
			\beta \times \left(  \pm \dot{B}\sigma + \dot{C} \right)	= 4 B^2
			\]
		This implies $B = constant$ and thus $f = \pm B\sigma + C(\lambda)$. This is just the variable separable case considered earlier.

		Similarly, for $f=f_2$ we have --
			\[
			\beta \left[  \dot{A} + \tanh\left(  B\sigma + C \right) \times \left(  \dot{B}\sigma + \dot{C} \right) \right] = B^2 \left[  1 + 3 \tanh^2\left(  B\sigma + C \right) \right]
			\]
		Since, both sides represent functions analytic in $\sigma$, we compare their Taylor series around the point $\sigma = 0$. The comparison of first 5 Taylor coefficients gives --
			\[
			\dot{B} = \frac{3 B^3}{\beta} sech^4C \hspace{10pt};\hspace{10pt} \dot{C} = \frac{3B^2}{\beta} \tanh C \left(  1 + \tanh^2C \right)
			\]
			\[
			- B^2 + \dot{A}\beta + 3B^2\tanh^4C = 0
			\]
			\[
			4 B^5 \tanh C sech^4C =0	
			\]
			\[
			B^6 \left(  4 -3 \cosh 2C \right) sech^6C = 0
			\]

		The only solution to these is $B = 0$ i.e. $f = A(\lambda) + \ln\left[ \cosh C(\lambda) \right]$ and again $f$ is separable. This is a limiting case of the earlier solution when $\lambda_c \rightarrow \infty$ and leads to a static flat metric.

		Thus, Eq.\ref{metric soln} is the general (and only) 
solution of the Ricci flow equations, when $r_c = r_c(\lambda)$.

	\subsection{Nonseparable metric functions: numerical results}

To deal with a more general situation, we now drop the assumption, 
$r_c = r_c(\lambda)$ and consider the full non--separable 
case of Eqs.\ref{f flow} \& \ref{r flow}. We observe that if 
$f(\sigma,\lambda)$ and $r_c(\sigma, \lambda)$ are a solution of the flow 
then so are $f(\alpha\sigma, \alpha^2\lambda)$ and $r_c(\alpha\sigma, \alpha^2\lambda)$. Thus, we look for \emph{scaling solutions} of the form $f(y)$ and 
$r_c(y)$ with $y=\frac{\sqrt{\lambda}}{\sigma}$. Note that it is also
possible to work with $\frac{1}{y}$ instead of $y$, though, with such a
choice, large $y$ will correspond to small $\lambda$ and vice versa.   
We further assume that
either $\sigma>0$ or $\sigma_0 <\sigma<\sigma_1$, i.e. we consider, as the
extra dimensional space the open half line or an open interval.  

		In terms of the new variable $y$ and defining
$u=\frac{1}{r_c}$,  the flow equations reduce 
to the following first order dynamical system,
			\be\label{f dyn sys}
			\frac{df}{dy} = A		
			\ee
			\be\label{A dyn sys}
			\frac{dA}{dy} = \frac{A }{2y^3 u^2} - \frac{2A}{y} 
- 24 y^3 A^3 u^2		
			\ee
			\be\label{r dyn sys}
			\frac{du}{dy} = -4 A u^3 \left( \frac{1}{u^2} - 
{6} A y^3 \right )
			\ee		
		Also, from Eq.\ref{ricci scalar} we can write --

			\be\label{ricci scalar y}
				\begin{split}
					\sigma^2 R & = -\frac{8y^2}{r_c^2}\left[ \frac{d^2f}{dy^2} + \frac{2}{y} \left(\frac{df}{dy}\right) + \frac{5}{2} \left(\frac{df}{dy}\right)^2 - \frac{1}{r_c} \left(\frac{df}{dy}\right)\left(\frac{dr_c}{dy}\right) \right]\\
						   & = -\frac{4A}{y} + 12y^2 A^2u^2
			 	\end{split}
			\ee
We shall use this latter expression in our evaluations.

Note that the above dynamical system is invariant under scaling transformations of the type $\lb( y \rightarrow \alpha y , f \rightarrow f , A \rightarrow A/\alpha , r_c \rightarrow \alpha r_c \rb)$. We can, thus, choose the scale by specifying initial conditions at $y = 1$ and solutions for conditions at other values of $y$ can be obtained simply by the appropriate scaling. 
Hence, starting from $y = 1$ we can study the evolution of the system in 
the future ($y>1$).
	
\begin{figure} 
\centering
\subfigure[$f$ v/s $y$]
{
\includegraphics[width=0.4\textwidth]{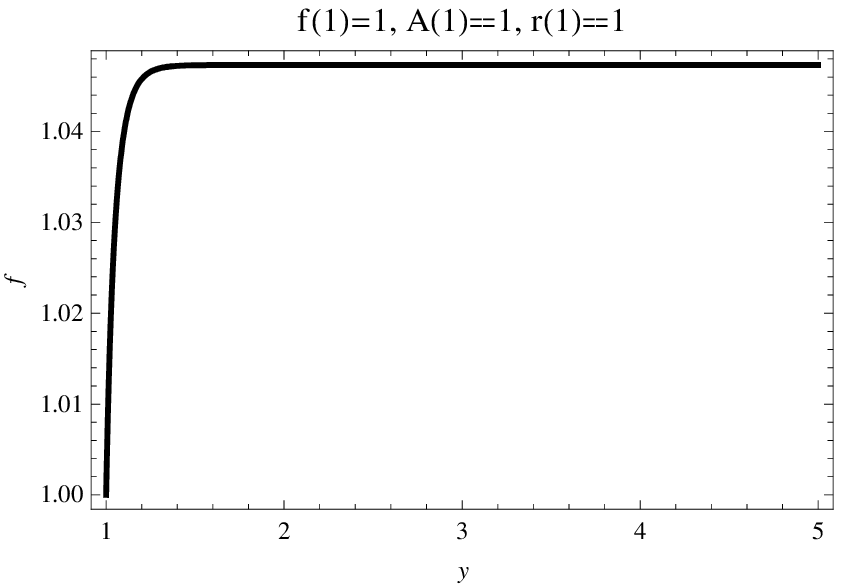} 
\includegraphics[width=0.4\textwidth]{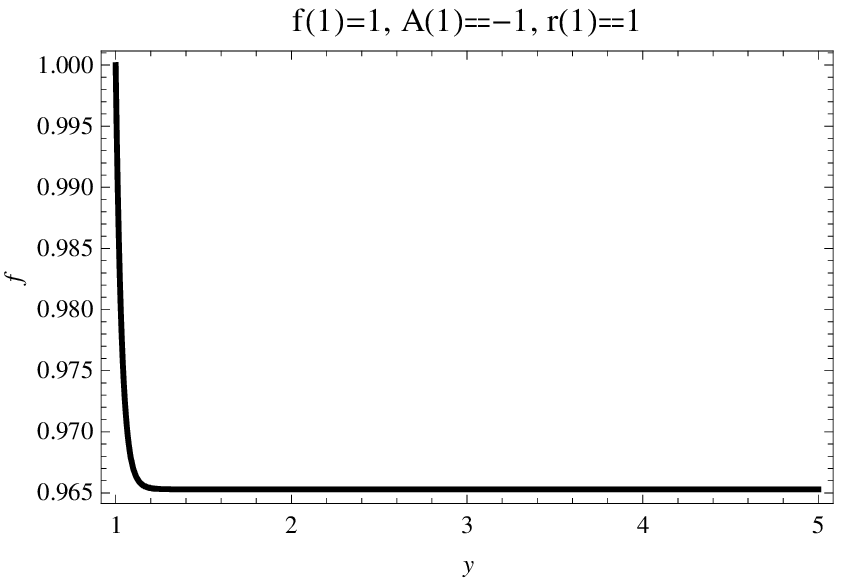} 
\label{fig fsoln}
}
\subfigure[$A$ v/s $y$]
{
\includegraphics[width=0.4\textwidth]{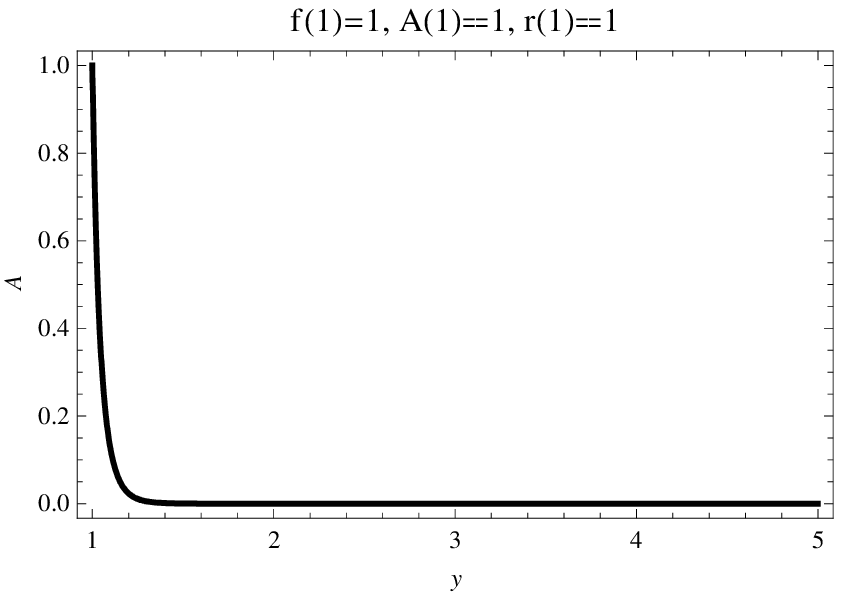} 
\includegraphics[width=0.4\textwidth]{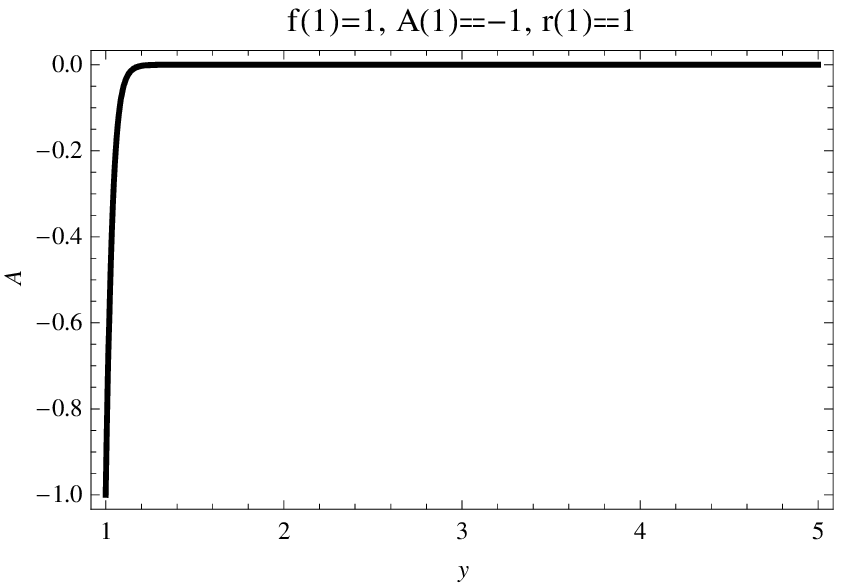}
\label{fig Asoln}
}
\subfigure[ $r_c$ v/s $y$]
{
\includegraphics[width=0.4\textwidth]{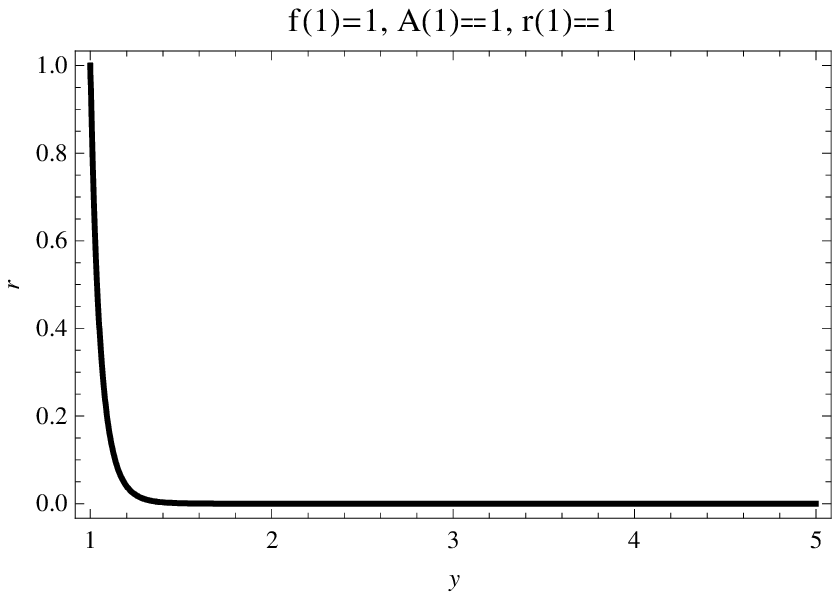}
\includegraphics[width=0.4\textwidth]{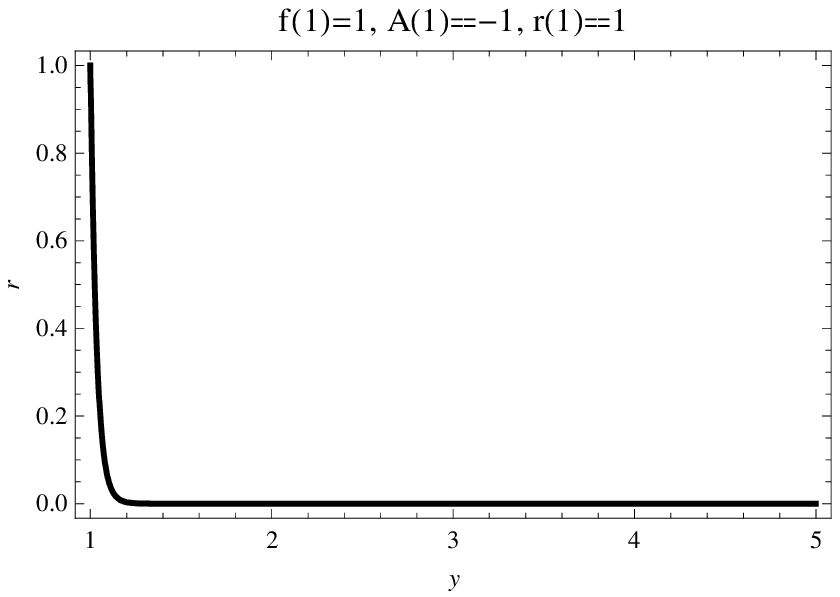} 
\label{fig rcsoln}
}
\subfigure[$\sigma^2R$ v/s $y$]
{
\includegraphics[width=0.4\textwidth]{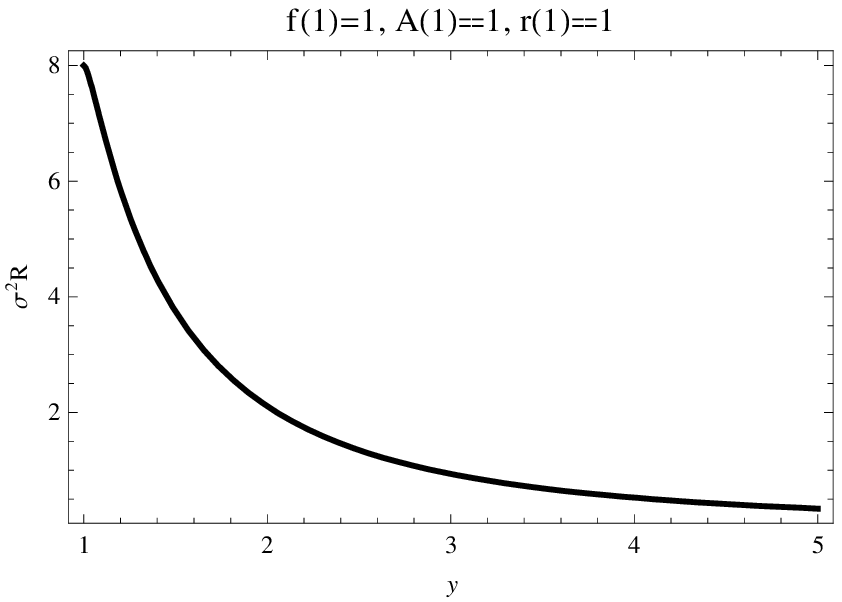}
\includegraphics[width=0.4\textwidth]{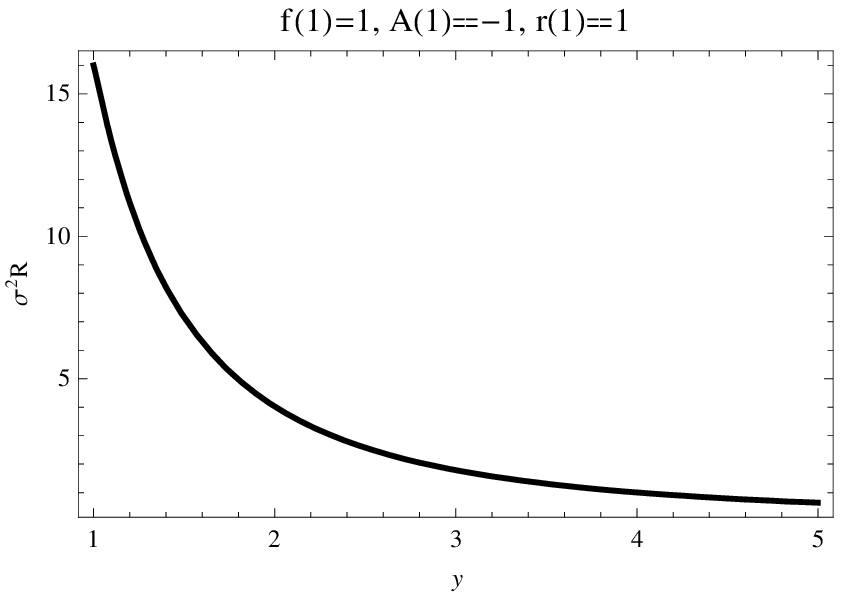} 
\label{fig Rsoln}
}
\caption{Results of numerical integration of dynamical system. Figures on left are for $A(1) = +1$ and those on right are for $A(1) = -1$}
\end{figure}

Further, as all our results will be in terms of $y =\frac{\sqrt{\lambda}}{\sigma}$ we must therefore fix $\sigma$ (at some $\sigma\neq 0$)
in order to understand the evolution in $\lambda$ at that $\sigma$ and 
subsequently repeat things for another $\sigma$. Physical considerations dictate that $r_c > 0$ and since the 
equations are independent of $f$ we can take $f(1) = 1$. We first choose the initial conditions at $y=1$ as $(f,A,r_c) = (1,\pm1,1)$. The results of the numerical 
integration are plotted in Fig.\ref{fig fsoln} - \ref{fig Rsoln}.\\

We note that when $A(1) = +1$, $f$ eventually goes to a constant and 
$r_c$ drops to zero in the future ($y>1$). 
Thus the flow exists for arbitrarily large values of $y$ i.e. in the future and the geometry tends towards a singular
geometry only as $y$ (or $\lambda$) becomes infinitely large. 

For $A(1) = -1$, as $y$ becomes large $f$ decreases and tends to a constant 
value, while
$A$ beginning at a negative value approaches zero eventually. $r_c$
tend to zero value as $y \rightarrow \infty$. The Ricci scalar
$R$ for both initial conditions on $A$ is non-singular at all finite $y$ and
decays with $y$ monotonically.\\

\begin{figure}
\centering
\includegraphics[width=0.4\textwidth]{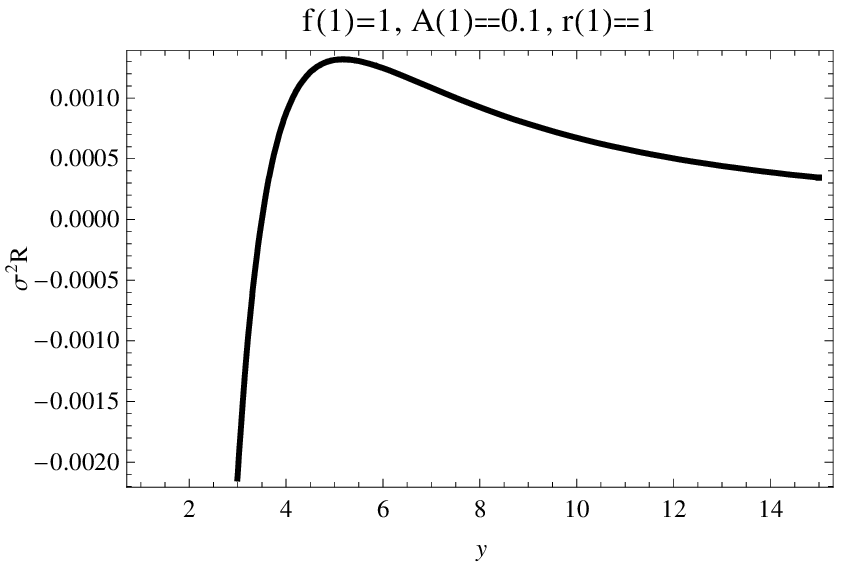}
\includegraphics[width=0.4\textwidth]{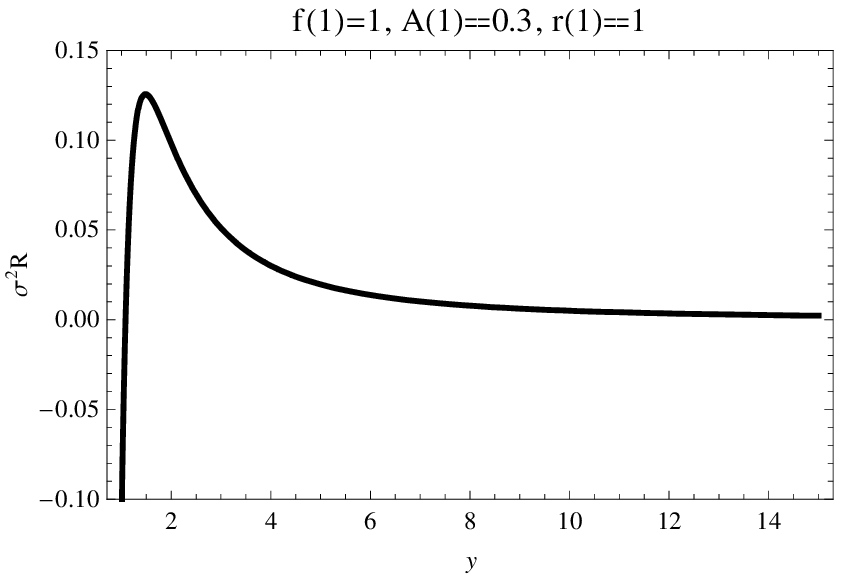}
\includegraphics[width=0.4\textwidth]{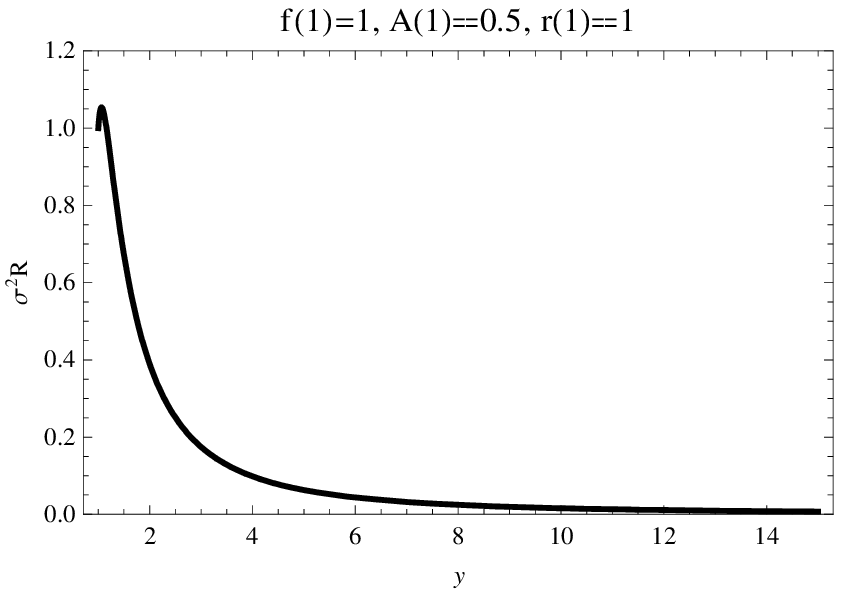}
\includegraphics[width=0.4\textwidth]{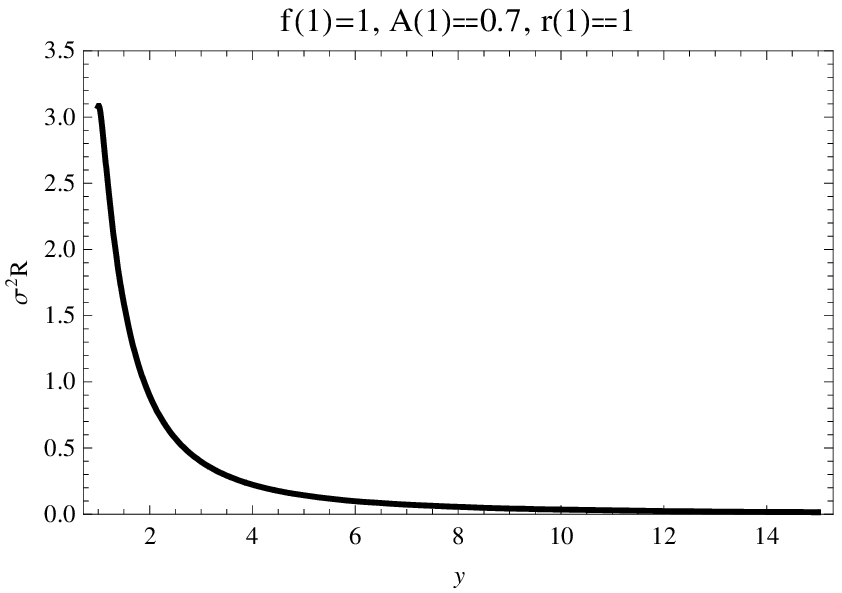}
\includegraphics[width=0.4\textwidth]{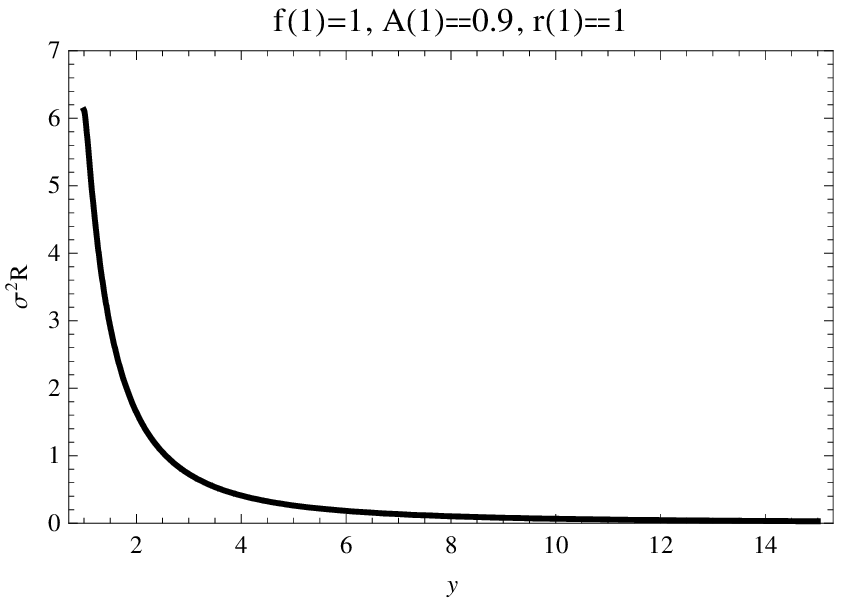}
\includegraphics[width=0.4\textwidth]{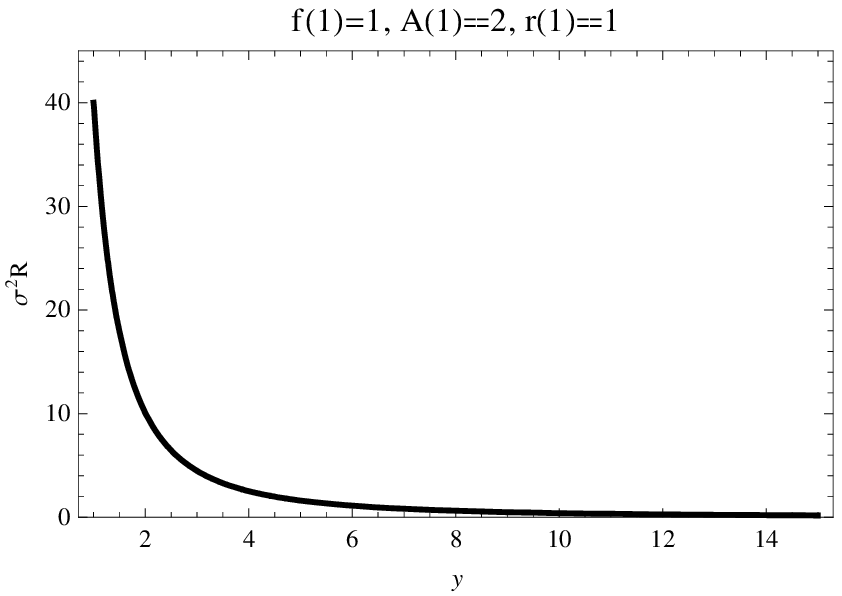}

\caption{$\sigma^2 R$ as a function of $y$ for fixed $f$, $r_c$ and changing $A$.}
\label{R varying A}
\end{figure}

It is necessary to look into the behaviour of the quantity $\sigma^2 R$
as we change the initial values of $f$, $A$ and $r_c$. Firstly, let us note 
that changing the initial value of $f$ does not produce any effect.
However, interesting features do arise when we change $A(1)$ or $r_c(1)$.
Fig.\ref{R varying A} shows this variation for different initial values of $A$. Starting from the top
left, we change $A(1)$ from a small value to successive larger values,
while keeping $r_c(1)$ and $f(1)$ fixed. We note that the location of
the maximum shifts towards vertical axis and the value of $\sigma^2 R$
at a given $y$ becomes larger with increasing initial $A$. Also, it is
worth noting that the $\sigma^2 R$ is always positive and goes to
zero asymptotically.

\begin{figure}
\centering
\includegraphics[width=0.4\textwidth]{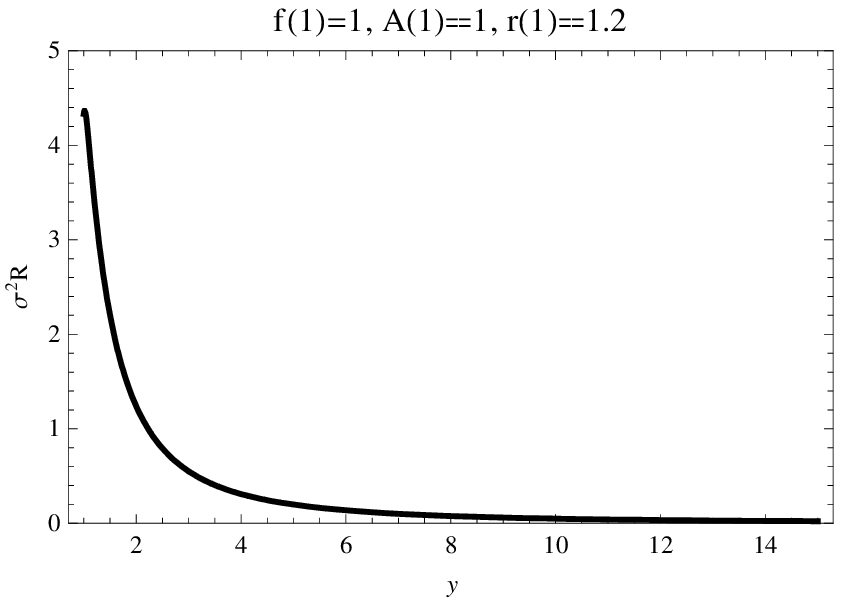}
\includegraphics[width=0.4\textwidth]{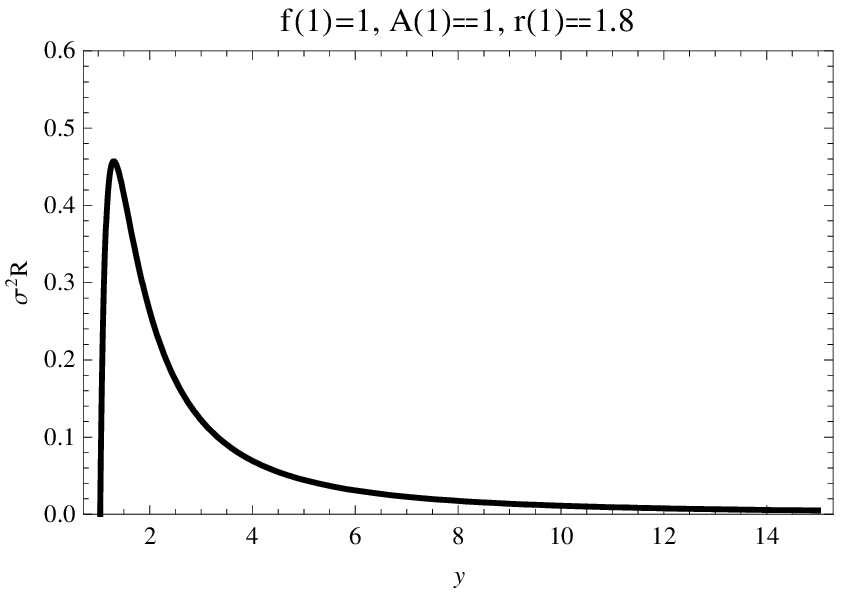}
\includegraphics[width=0.4\textwidth]{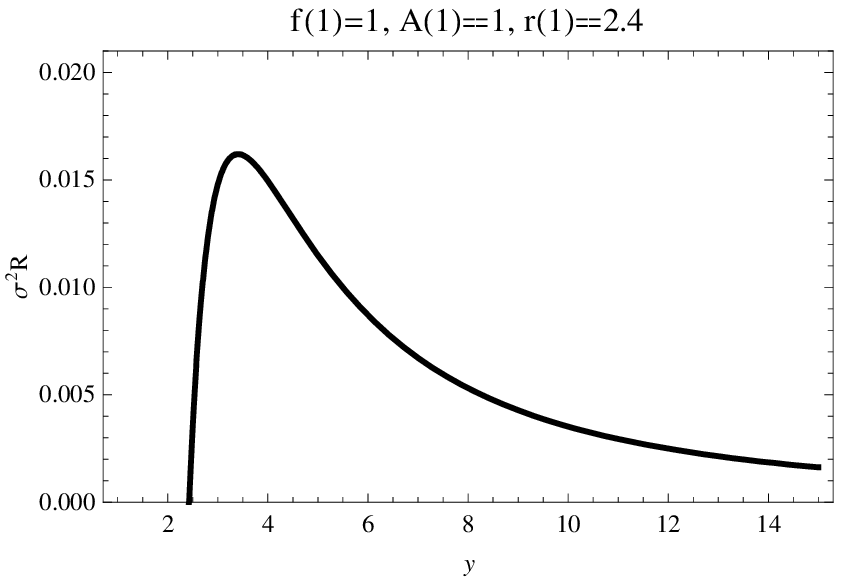}
\includegraphics[width=0.4\textwidth]{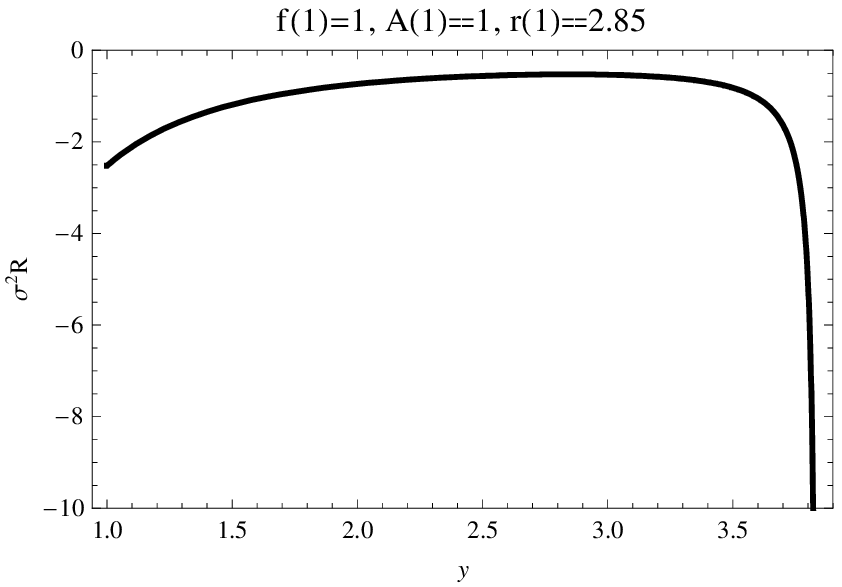}

\caption{$\sigma^2 R$ as a function of $y$ for fixed $f$, $A$ and changing $r_c$.}
\label{R varying rc}
\end{figure}

In Fig.\ref{R varying rc}, we show how a variation of $r_c(1)$ (at fixed $f(1)$ and $A(1)$) changes 
the behaviour of $\sigma^2 R$. At smaller values of $r_c(1)$  the flow 
continues indefinitely into the future. But for a critical value of $r_c(1)$, the flow develops a \emph{future singularity}. Moreover, with changing
$r_c$ we switch from positive to negative values of $\sigma^2 R$. 
The flow cannot be extended beyond a certain value of $y$
if the initial $r_c$ is above a certain critical value. \\

\begin{figure}
\subfigure[future singularities]{\includegraphics[width = 0.4\textwidth]{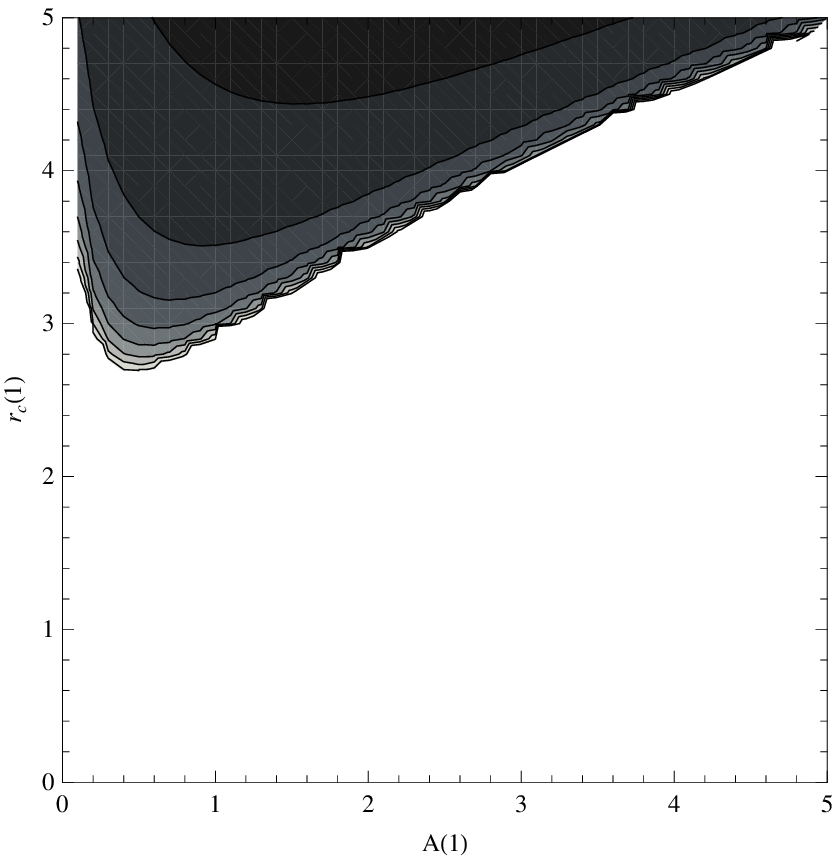}}
\caption{phase diagram showing non-singular and singular flows in space of ($A(1)$, $r_c(1)$)}
\label{phase diagram}
\end{figure}

To understand the appearance of genuine singularities, 
we investigate when such future singularities arise in the flow 
and their 
dependence on the initial conditions on $r_c$ and $A$,. 
We search the range of $1 < y <100$ 
with different initial conditions on $A$ and $r_c$, 
and look for the appearance of singularities. Since, the 
flow does not depend on $f$, we can make a \emph{phase diagram} in the 
space of ($A(1), r_c(1)$), in which we can represent 
singular and non-singular flows. 
Fig.\ref{phase diagram} shows such a plot. 
The white (blank) region represents conditions that do not lead to 
singularities, whereas the darker shades of gray denote flows with 
value of $y = y_s$ at which the flow becomes singular. Thus, for future singularities, increasing $r_c(1)$ makes $y_s$ smaller, and hence the flow can only be 
continued to shorter times in the future. These trends can also be seen in 
Figs.\ref{R varying A} and \ref{R varying rc}. Note that for $A(1) < 0$ 
the flow has no singularities at all and exists for all time. 
Also, from 
the phase diagram in Fig.\ref{phase diagram} we see that, there are 
regions of initial conditions $A(1), r_c(1)$ corresponding to 
both types of flow --- non-singular and only future singular.\\

Given the above results, let us now try to understand the consequences
a little more. From the solutions for $f$, $A$, $r_c$ and the expression
for $\sigma^2 R$
(all of which are functions of $y$) we note that for fixed $y=y_0$,
all these quantities have a fixed value, i.e. $f(y_0)=f_0$, $A(y_0)=A_0$,
$r_c (y_0)= r_{co}$ and $\tilde{R}(y_0) =\sigma^2 R (y_0)= \tilde{R}_0$.
However, we also have $y_0=\frac{\sqrt{\lambda}}{\sigma}$ or $\sigma=\frac{\sqrt{\lambda}}{y_0}$. Hence. for fixed $y$, 
the flow along $\lambda$ is equivalent to moving along the extra dimension. 
Changing $\sigma$ (or $\lambda$) keeps the obtained $f$, $A$, $r_c$ and
$\sigma^2 R$ unchanged at fixed $y$. However, $R$ will change with changing
$\sigma$ and at different $\sigma$ the value of the Ricci scalar will
obviously be different. Thus, the solutions obtained represents 
Ricci flow along the extra dimensional
coordinate (proportional to $\sqrt{\lambda}$ at fixed $y$),
such that the value of $f$, $A$ and $r_c$, as we move from
one location to another in
the extra dimension remains unaltered, though the value of the five 
dimensional Ricci scalar changes.  

\section{Remarks and conclusions}
We conclude with a summary and some remarks.

We have looked into Ricci flow of unwarped and warped product manifolds.
In the unwarped case we have focused on $\mathbb S^n\times \mathbb S^m$, $\mathbb
H^n \times \mathbb H^m$ and $\mathbb S^n\times \mathbb H^m$. Distinguishing
features of the Ricci flow of such manifolds have been highlighted with
reference to fixed points/curves, singularity formation, intersections
which imply isotropisation at some $\lambda$ and dependence on initial
conditions. We have also discussed
multiple products and provided illustrative examples for the case
$\mathbb S^p \times \mathbb S^n \times \mathbb S^m$.

Further, inspired by warped braneworlds, we have looked into Ricci
flow of such manifolds with arbitrary functional forms of the 
warp factor and the scale of the extra dimensions. We discuss
two separate cases--one where these functions are separable
(this being analytically solvable) and the other, the non--separable
case where we use numerical methods to find the evolution. In the
former case, we are able to find the line element exactly and also
compute its curvature. We note that the spacetime is necessarily
AdS for all $\lambda$ and this $\lambda$ can, in some way, be
associated with the $k$ (AdS curvature scale) in the warped braneworld 
scenario. Thus
Ricci flow seems to lead to a RS type warp factor and the
curvature of the spacetime remains AdS during evolution.
In the more general case, we have analysed numerically the
particular case of scaling solutions.
We find that, for certain initial conditions, 
the evolution is regular throughout.
We have shown that Ricci flow in $\lambda$ turns out to
be equivalent to a flow along $\sigma$ (at fixed $y$)
and the $f$, $A$,$r_c$ and $\sigma^2 R$ at that value of $y$ 
do not change with varying $\lambda$ (or $\sigma$), though the
value of $R$ does evolve. For other initial conditions, we
note that genuine singularities may appear at finite $y$.
We also observe a transition from non-singular to singular solutions as we tune the initial conditions. 
 
Finally, we mention that we hope to look into more complex and
realistic scenarios in the warped context, in future. In the
unwarped case, we would like to arrive at more general conclusions
for arbitrary multiple products by a more detailed analysis of the 
associated dynamical system.


\end{document}